\documentclass[11pt]{article}

\usepackage{times}
\usepackage{epsf}

\newenvironment{sciabstract}{%
\begin{quote} \bf}
{\end{quote}}

\newcounter{lastnote}

\title{Mapping the Large Scale Structure of the 
Universe\footnote{Appeared in {\it Science}, Vol. 309, p. 564 (2005)}}

\author
{David H. Weinberg\\ 
\\
\normalsize{Department of Astronomy, Ohio State University,}\\ 
\normalsize{140 W. 18th Ave., Columbus, OH 43210, USA}\\
\normalsize{E-mail:  dhw@astronomy.ohio-state.edu}
}

\date{}

\begin{document} 




\maketitle 


\begin{sciabstract}
\end{sciabstract}

In a large scale view of the universe, galaxies are the basic unit of
structure.  A typical bright galaxy may contain 100 billion stars and
span tens of thousands of light years, but the empty expanses between the
galaxies are much larger still.  Galaxies are not randomly distributed in 
space, but instead reside in groups and clusters, which are themselves 
arranged in an intricate lattice of filaments and walls, threaded by tunnels
and pocked with bubbles.  
Two ambitious new surveys, the Two-Degree Field Galaxy Redshift Survey
(2dFGRS) and the Sloan Digital Sky Survey (SDSS), have mapped the
three-dimensional distribution of galaxies over an unprecedented
range of scales ({\it 1, 2}).  Astronomers are using these maps
to learn about conditions
in the early universe, the matter and energy contents of the cosmos,
and the physics of galaxy formation.

Galaxies and large scale structure form by the gravitational
amplification of tiny primordial fluctuations in the density of
matter.  The inflation hypothesis ascribes the origin of these
fluctuations to quantum processes during
a period of exponential expansion that occupied the first 
millionth-of-a-billionth-of-a-trillionth of a second of
cosmic history.  Experiments over the last decade have revealed
the imprint of these fluctuations as part-in-100,000 intensity
modulations of the cosmic microwave background (CMB), 
which records the small inhomogeneities present in the universe
half a million years after the big bang.  
While the visible components of galaxies are made of ``normal''
baryonic matter (mostly hydrogen and helium), the
gravitational forces that drive the
growth of structure come mainly from dark matter, which
is immune to electromagnetic interactions.

By combining precise,
quantitative measurements of present-day galaxy clustering with
CMB data and other cosmological observations, astronomers
hope to test the inflation hypothesis, to 
pin down the physical mechanisms of inflation,
to measure the amounts of baryonic and dark matter in the cosmos,
and to probe the nature of the mysterious ``dark energy'' that has
caused the expansion of the universe to accelerate over the last 
5 billion years.  
The 2dFGRS, completed in 2003, measured distances to 220,000 galaxies,
and the SDSS is now 80\% of the way to its goal of 800,000 galaxies
(see figure).  The key challenge in interpreting the observed clustering
is the uncertain relation between the distribution of galaxies and 
the underlying distribution of dark matter.
If the galaxy maps are smoothed over tens of millions of light years,
this relation is expected to be fairly simple:
variations in galaxy density are constant multiples of the variations
in dark matter density.  Quantitative analysis in this regime has
focused on the power spectrum,
which characterizes the strength of clustering on different scales
({\it 3, 4}).
In geographic terms, the power spectrum 
describes the contributions of mountain ranges,
isolated peaks, and rolling hills to the cosmic landscape.
The shape of the dark matter power spectrum is a 
diagnostic of the inflation model, which predicts the input spectrum
from the early universe, and of the average dark matter density, which 
controls the subsequent gravitational growth.  
Recent analyses have also detected subtle modulations
of the power spectrum caused by baryonic matter, which undergoes
acoustic oscillations in the early universe because of its interaction
with photons ({\it 4, 5}).

To go further, one would like to know the precise amplitude of 
dark matter clustering, not just the variation of clustering with
scale.  Unfortunately, the factor relating galaxy and dark matter
densities depends on aspects of galaxy formation that
are difficult to model theoretically.  One observational approach
to isolating dark matter clustering uses weak gravitational lensing,
in which the dark matter surrounding nearby galaxies subtly distorts
the apparent shapes of the distant galaxies behind them.
Another approach uses the probabilities of different 
configurations of galaxy triples to pick out the characteristic,
elongated-triangle signature of anisotropic gravitational collapse.
Applications of these two methods
to the SDSS and the 2dFGRS, respectively, imply that the clustering
strength of bright, Milky Way-type galaxies is similar to that of 
the underlying dark matter ({\it 6, 7}).

On smaller scales, the relation between galaxies and dark matter 
becomes more complex, and it is different for different types of galaxies.
Redder galaxies composed of older stars reside primarily in clusters, 
the dense urban cores of the galaxy distribution.
Younger, bluer galaxies populate the sprawling,
filamentary suburbs.
Current efforts to model galaxy clustering in this regime
focus on the ``halo occupation distribution,'' a statistical
description of the galaxy populations of gravitationally bound
``halos'' of dark matter.
Depending on its mass, an individual
dark halo may play host to a single bright galaxy, a small group of 
galaxies, or a rich cluster.  

By combining theoretical predictions for the 
masses and clustering of halos with precise measurements
of the clustering of galaxies,
one can infer the halo occupation distribution 
for different classes of galaxies empirically.
Theoretical models of galaxy formation predict a strong dependence
of halo occupation on galaxy luminosity and color, and the initial
results from the 2dFGRS and the SDSS show good qualitative agreement
with these predictions ({\it 8, 9}).
Increased precision and measurements for more galaxy classes
will test the predictions in much greater detail, and they will sharpen
our understanding of the physical processes that produce
visible galaxies in the first place and determine their observable properties.
By deriving the relation between galaxies and dark matter from
the clustering data themselves, halo occupation methods also 
allow new cosmological model tests that take advantage of precise
measurements on small and intermediate scales.

The large scale clustering results from the 2dFGRS and the SDSS,
in combination with CMB measurements and other cosmological data,
support the predictions of a simple inflation model in a universe
that contains 5\% normal matter, 25\% dark matter, and 70\% dark energy
({\it 10}).  However, several analyses that incorporate 
smaller scale clustering
suggest that either the matter density or the
matter clustering amplitude is lower than this ``concordance'' model
predicts, by 30-50\% ({\it 11, 12, 13}).
This tension could reflect
systematic errors in the measurements or the modeling, but it could
also signal some departure from the simplest models of primordial
fluctuations or dark energy.
For example, if inflation produces gravity waves that contribute
to observed CMB fluctuations, then
naive extrapolation of these fluctuations would overpredict the
level of matter clustering today.
Alternatively, evolution of the dark energy component can affect the amount
of growth since the CMB epoch.
As the SDSS moves towards completion, 
improved clustering measurements and analyses may restore the consensus
on a ``vanilla'' cosmological model, or they may provide sharper evidence
that our theoretical recipe for the universe is still missing
a key ingredient.

\subsection*{References}
\begin{itemize}
\item[1.]
M. Colless {\it et al.}, 
{\it Mon. Not. Roy. Ast. Soc.}, {\bf 328}, 1039 (2001)
\item[2.]
D. G. York {\it et al.},
{\it Astron. J.}, {\bf 120}, 1579 (2000)
\item[3.]
M. Tegmark {\it et al.},
{\it Astrophys. J.}, {\bf 606}, 702 (2004)
\item[4.]
S. Cole {\it et al.},
{\it Mon. Not. Roy. Ast. Soc.}, submitted,
preprint astro-ph/0501174 
\item[5.]
D. J. Eisenstein {\it et al.},
{\it Astrophys. J.}, submitted,
preprint astro-ph/0501171
\item[6.]
E. Sheldon {\it et al.},
{\it Astron. J.}, {\bf 127}, 2544 (2004)
\item[7.]
L. Verde {\it et al.},
{\it Mon. Not. Roy. Ast. Soc.}, {\bf 335}, 432 (2002)
\item[8.]
F. C. van den Bosch, X. Yang, H. J. Mo,
{\it Mon. Not. Roy. Ast. Soc.}, {\bf 340}, 771 (2003)
\item[9.]
I.\ Zehavi {\it et al.}, 
{\it Astrophys. J.}, in press, 
preprint astro-ph/0408569
\item[10.]
Convergence on this model from several independent lines of
evidence was the 2003 {\it Science}
``Breakthrough of the Year'';
C. Seife, {\it Science}, {\bf 302}, 2038 (2003)
\item[11.]
F. C. van den Bosch, H. J. Mo, X. Yang,
{\it Mon. Not. Roy. Ast. Soc.}, {\bf 345}, 923 (2003)
\item[12.]
N. A. Bahcall {\it et al.},
{\it Astrophys. J.}, {\bf 585}, 182 (2003)
\item[13.]
J. Tinker, D. H. Weinberg, Z. Zheng, I. Zehavi,
{\it Astrophys. J.}, in press, 
preprint astro-ph/0411777
\end{itemize}

\begin{figure}
\centerline{
\epsfxsize=6.0truein
\epsfbox[15 210 590 580]{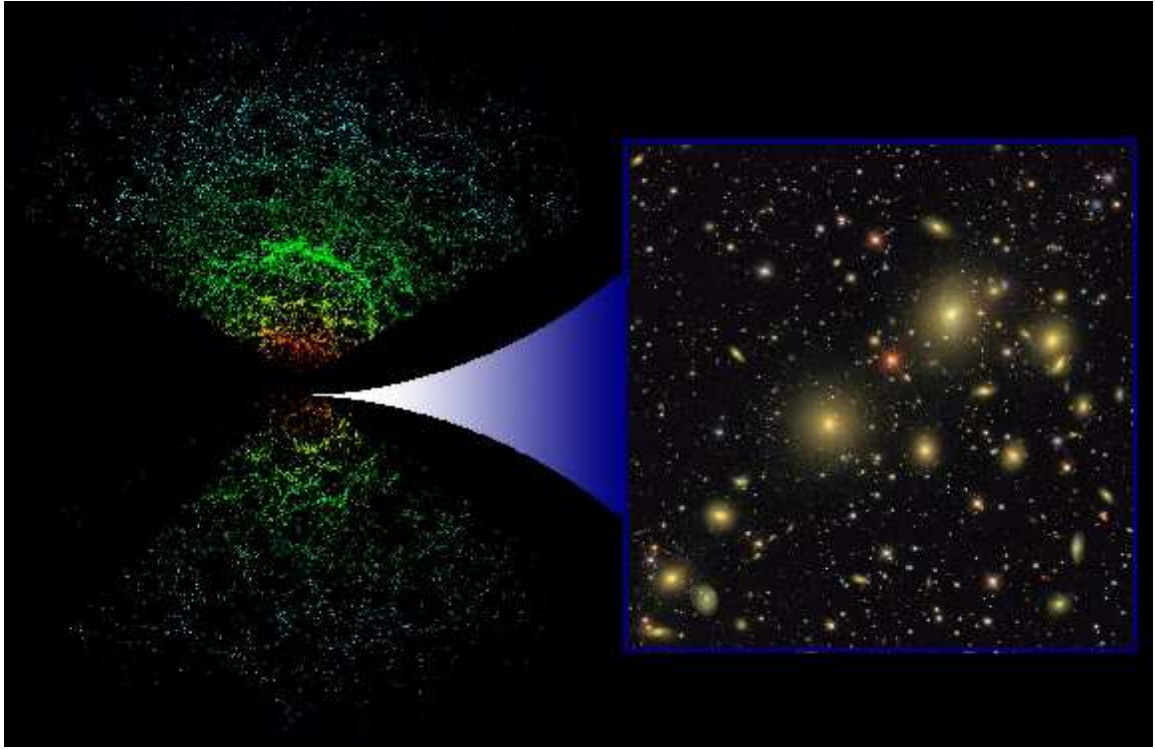}
}
\caption { 
Large scale structure in the Sloan Digital Sky Survey (SDSS).
The SDSS uses a mosaic CCD camera to image regions of the sky
and a fiber-fed spectrograph to measure distances of galaxies 
selected from these images.  The main panel shows the SDSS
map of 67,000 galaxies that lie within 5$^\circ$ of the equatorial
plane; the region of sky obscured by the Milky Way is not mapped.
Each wedge is 2 billion light years in extent.  Galaxies are
color-coded by luminosity, and more luminous galaxies can be seen
to greater distances.  {\bf Inset:} SDSS image of a cluster
of galaxies, showing a region roughly 1 million light years on a side.
}
\label{fig:sdss}
\end{figure}

\end{document}